\documentclass[conference,landscape]{IEEEtran}
\IEEEoverridecommandlockouts
\usepackage{cite}
\usepackage{amsmath,amssymb,amsfonts}
\usepackage{algorithmic}
\usepackage{graphicx}
\usepackage{textcomp}
\usepackage{xcolor}
\usepackage{multirow}
\usepackage{multicol}
\def\BibTeX{{\rm B\kern-.05em{\sc i\kern-.025em b}\kern-.08em
    T\kern-.1667em\lower.7ex\hbox{E}\kern-.125emX}}
\begin{document}

\title{Real time Smart Contracts for IoT using Blockchain and  Collaborative Intelligence based Dynamic Pricing for  the next generation Smart Toll Application\\
}

\author{\IEEEauthorblockN{ Misha Abraham}
\IEEEauthorblockA{\textit{Robert Bosch Engineering and } \\
\textit{Business Solutions Private  Limited}\\
Bangalore, India \\
Abraham.Misha@in.bosch.com}
\and
\IEEEauthorblockN{ Himajit Aithal}
\IEEEauthorblockA{\textit{Robert Bosch Engineering and } \\
\textit{Business Solutions Private  Limited}\\
Bangalore, India \\
Himajit.Aithal@in.bosch.com}
\and
\IEEEauthorblockN{ Krishnan Mohan}
\IEEEauthorblockA{\textit{Robert Bosch Engineering and } \\
\textit{Business Solutions Private  Limited}\\
Bangalore, India \\
Krishnan.Mohan@bosch.com
}

}

\maketitle

\begin{abstract}

The confluence of Internet of Things(IoT) , Blockchain(BC) and Artificial Intelligence(AI) acts as a key accelerator for enabling Machine Economy \cite{m00}. To be ready for  future businesses these technologies needs to be adapted by extending the IoT capabilities to Economy of Things (EoT) \cite{m0} capabilities. In this paper we focus on one such implementation experience for Smart Toll Transaction application in the domain of mobility. Our paper showcases a possible solution  by leveraging negotiations, decision making, distributed learning capabilities at the devices level using AI-enabled Multi-Agent Systems and the real-time smart contracts between the Cars and Tolls using Blockchain. This solution also showcases the monetization of real time data coming from various IoT devices which are part of vehicles and infrastructure. While blockchain secures the privacy of the participants it also acts as an economic transactional layer and governance layer between the devices in the network.

\end{abstract}


\section{Introduction}
Increasing computation power as per Moore's Law and increasing connectivity as per Nielsen's Law, are enabling the IoT-edge devices to run software entities that represent and mimic human decision making and negotiations in the digital domain. The connected IoT networks shall have different software entities also known as agents which acts autonomously and learns from distributed sources, to maximize their owner's interests in a sustainable, fair and transparent manner. Two new emerging trends have led to renewed interest in enhancing the field of internet of things: \\ 
-Blockchain/Distributed ledger technology : Enabling robust, self-governing economic transnational systems, by decentralizing control layers across the network (nodes, edge devices etc.,), also preserving privacy and confidentiality.\\
- Multi agent Systems: Self executable software entities that can learn, take decisions and maximize the incentives based on configurable business rules and collaborative learning.\\

The Internet of Things (IoT) is fascinating and exciting topic, but the major difficulty in this field is to have a secure environment comprising all the building stones \cite{m1}. The  devices in IoT are connected and in today's world, they lack authenticity, security, privacy and also economic transactions. In this paper, we focus on bringing in the idea of blockchain technology and multi-agent systems to fill the gaps in the IoT ecosystem. The "Smart Toll Transaction" as a business case was picked up, as we see lots of relevance on the applicability of new-age technologies (Blockchain, IoT and AI) to bring changes in today's toll transactions. We could verify that what is taking days/weeks for the settlement in toll transactions can be bought down to instant settlements.

We envision a hybrid approach comprising of Global (Long Term) optimization and Local (Short Term) optimization. Global optimizations look at multiple dimensions of the system and entities to predict the best possible approach an agent should take to maximize its gains. One of the dimensions could be historical data which a user considers non-sensitive and shareable. Local optimizations are more reactive in nature, where localized (sensory and agent critical data) data and contextualized (data shared by neighbouring trusted agents) data are used to make the final business decisions. In short, global optimizations provide a heuristic strategy model for the agents, while local optimizations fine-tune the models based on factors such as preferences of their users, current ecosystem conditions, etc. Thus each agent exists in a coopetitive (collaboration + competition) environment, where they compete on the global level but collaborate on the local level. To illustrate the concept, consider an ambulance with a critical patient on a busy highway. Global optimization provides the shortest path to the hospital. Local optimization talks to nearby vehicles and provides the right way to the ambulance. The ambulance agent has the intelligence to incentivize infrastructure like traffic signals, for changing the control actions of the infrastructure to reach its destination faster. 

The continuation of the paper is structured as follows.  Section II briefly explains the literature survey done on different areas like blockchain, multi-agent systems, blockchain in IoT, Blockchain and multi-agent systems, Smart Toll, etc., Section III explains the current issues and the scope of the problem in today's world. Section IV brings in the solution architecture of the problem and followed by the implementation details of the demonstrator in section V. Section VI explains the experimental analysis and future work. Section VII gives in a summary of the current work and the impact on the IoT ecosystem.

\section{Literature Survey}
\subsection{Blockchain}
Blockchain technology is proven to be one of the possible framework for developing a secure protocol for multi-agent business \cite{m2}. Blockchain technology makes it possible to create a peer-to-peer decentralized network maintaining integrity \cite{m3} and serves as a trusted communication mechanism for the devices and the multi-agent systems. This technology maintains an immutable and verifiable track  of records and avoids the need for a third party by achieving distributed consensus among the participants \cite{m20}.
\subsection{Blockchain for IoT}
Sharma et al. \cite{m4} introduced an SDN architecture (DistBlockNet) for IoT devices which is secure and distributed. The architecture proposes a peer-to-peer network which avoids the need for a trusted third party for communication among the members in the network. Also without human management, the security mechanism in this architecture can adapt automatically to the threat conditions. Also, the architecture performance was analyzed and compared. Daza et al. \cite{m5} proposed a methodology to discover the IoT using multi-layer blockchain technology. The method uses information obtained from each blockchain and existing service provider's protocol to detect the environment and discover the  network. 
\subsection{Multi-Agent Systems}
In a Multi-agent system, an agent is considered as an autonomous entity capable of observing the surrounding environment and interacting with it and also between other agents in the network \cite{m11}. It can be used to develop intelligent distributed network. Blockchain technology can be well utilized to bring faster-trusted interaction among the agents in multi-agent systems \cite{m22}. Also, Smart Contract, one of the key development that emerged as part of blockchain technology can be used between multi-agent systems to automate the business rules and transfer commands between agents \cite{m23}. The method allows us to write custom logic and storing the resulting data into the blockchain. Blockchain ensures secure and automated communication between agents. In IoT devices agents play a vital role and along with blockchain it brings in more values to the IoT ecosystem \cite{m24}. 
\subsection{Blockchain and Multi-agent systems}
In this section, we bring some of the publications where blockchain technology is applied in the field of multi-agent systems.
The paper \cite{m6} aims at bringing in a protocol using blockchain technology to overcome the issues faced in economic and industrial works due to the centralized management of agents. The protocol proposed a claim to overcome changing information and storage-related issues in a centralized agent  system. The authors have also shared their experience while implementing the protocol in a multi-agent system with unnamed aerial vehicles.
The paper \cite{m7} brings in a detailed literature review on the existing contributions that include multi-agent systems and blockchain technology. The authors have also justified on how blockchain technology addresses the multi-agent system requirements. The paper also covers the challenges of using blockchain technology in practice particularly in the field of multi-agent systems and ends with future research directions.
The paper \cite{m8} covers an overview of blockchain-based solutions for multi-agent robotic systems and also gives a classification and analysis of the technology. The paper also explains how the architectures, frameworks, and structures of blockchain technology can be used to solve the problems faced by the multi-agent robotic systems and cyber-physical systems.
The paper \cite{m9} presents a system including blockchain technology and multi-agent systems that aim at building a base for trusted interaction. The system also enables more trustworthy ways for an agent reputation computation. The authors have also brought in the arguments about the choice of applying Blockchain technology in the field of multi-agent systems and also their experience gained while testing the system in four scenarios with different configurations (normal execution, including down-agents or malicious agents).
\subsection{Smart Toll Applications Using Blockchain}
The patent \cite{ml1} focuses on implementing a data management system for Smart toll collection using the blockchain technology. The blocks in the blockchain stores the data and are encrypted and connected to the previous blocks which helps in maintaining the integrity of the data. The authors have given a well structured architecture diagram for their proposal and also proofs to state the correctness of the proposal. The invention \cite{ml2} provides a payment system for highway using blockchain technology which tries to solve problems like uncertainty in the judgement of vehicles, information of vehicles etc., and misuse of centrally stored data. The patent \cite{ml3} proposes an architecture for collecting traffic congestion fee using blockchain technology. The architecture does the following: report current location to the blockchain enabled node and based on the location decides whether the automobile passes a road segment with traffic congestion and if yes, traffic congestion fee is charged to the user. The users of the automobile act as nodes in the blockchain and have wallets with some initial balance. Money is deducted from the wallet when the user of the automobile is identified to be in a traffic congestion road. 

The papers and patents mentioned above focuses on exploring the applicability of blockchain technology and multi-agent systems in different areas but still lacks in defining the value proposition  for a given business use-case. In this paper, we focus on a particular use case, Smart toll transactions, and demonstrating how the convergence of IoT, blockchain and multi-agent systems can bring in new values like dynamic pricing, collaborative learning, instant settlement, micro payments etc., in a machine to machine economic scenario.


\section{Current Issues}
The toll transactions in today's world have fixed pricing for the lanes and also the users have no option to choose the lane based on their preference (fast lane or economic lane). Also modern-day toll systems take days/weeks for the settlement of toll transactions. As studied by the European Union parliament the cross border exchange and the lack of interoperability between the Toll systems are leading to a loss of millions of euros per year already \cite{m13}. This paper showcases a solution focusing on monetization of infrastructure, dynamic pricing for the lanes and instant settlements between car and tolls.

\section{Solution Architecture}
The concept of IoT has evolved from simple sensors, tasked to capture parametric data, to Economy of Things, where machines enter into negotiations and interact autonomously to perform tasks. However, even with advancements in such topics, today human decision making still forms the core part of any solution, from pricing models to the selection of trusted parties. Due to the high-frequency low-value nature of transactions in a machine-led economy, this decision making increases our cognitive load, making the process inefficient. We believe that the next step in this evolutionary process is Collaboration Among Things abbreviated as CaT, wherein each sensor or device is cooperating with trusted neighbouring sensors/devices in the network, which would help it to take a decision in a truly autonomous, secure and efficient manner. 
\begin{figure}[h]
\centerline{\includegraphics[width=80mm,height=60mm]{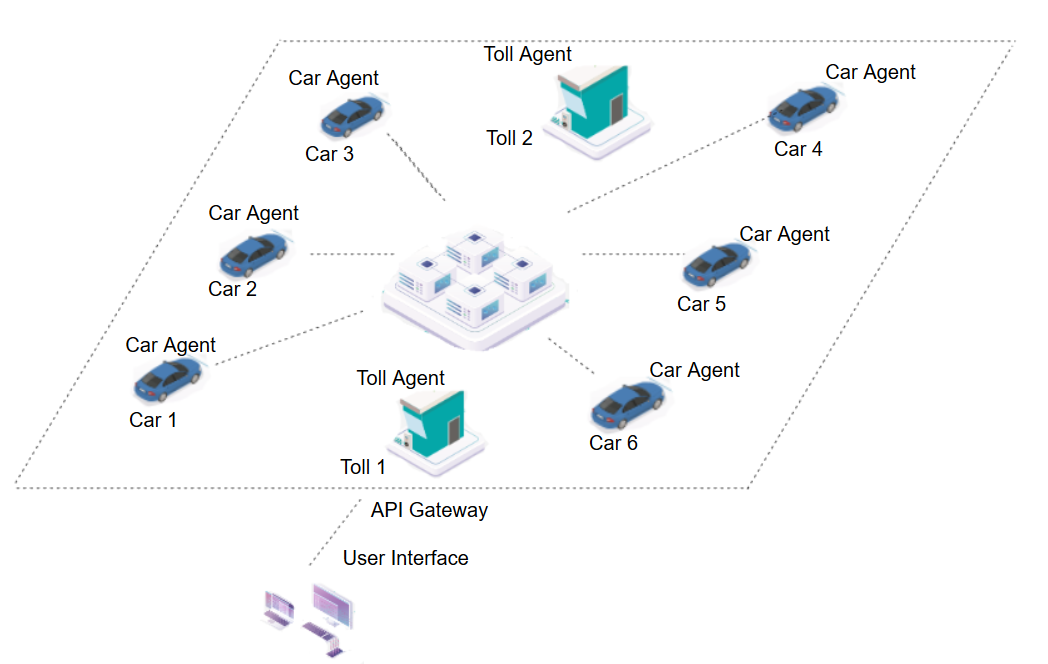}}
\caption{Proposed system Design for Smart Toll Transactions using blockchain technology and multi agent systems.}
\label{fig1}
\end{figure}

Figure \ref{fig1} shows the system design for Smart Toll Transactions. The transaction between the Toll and the Car has been captured in the immutable ledger and the car pays directly to the toll via cryptocurrency(Ethers). To interact with the blockchain network an interface is provided for the agent. The Toll agent learns from various sources like traffic density, from other toll, from user preferences and the historic transaction pricings, to determine what should be the current price of a lane thus demonstrating the Distributed Learning behavior. The cars and toll negotiate with each other in real-time for the lanes, finalize their economic transaction backed by cryptocurrencies and thus demonstrate Real-time contracting.
Blockchain-enabled multi-agent systems in Smart Toll Transactions benefits both the toll operator and the users. Dynamic Pricing, Enhanced Resource Optimization and  Instant Settlements are some of the benefits for the toll operators and Maximization of incentives (time/cost), Reduced wait times and Personalized preferences are some of the benefits of the users in the system.


\section{Implementation}
\begin{figure}[htbp]
\centerline{\includegraphics[width=80mm,height=60mm]{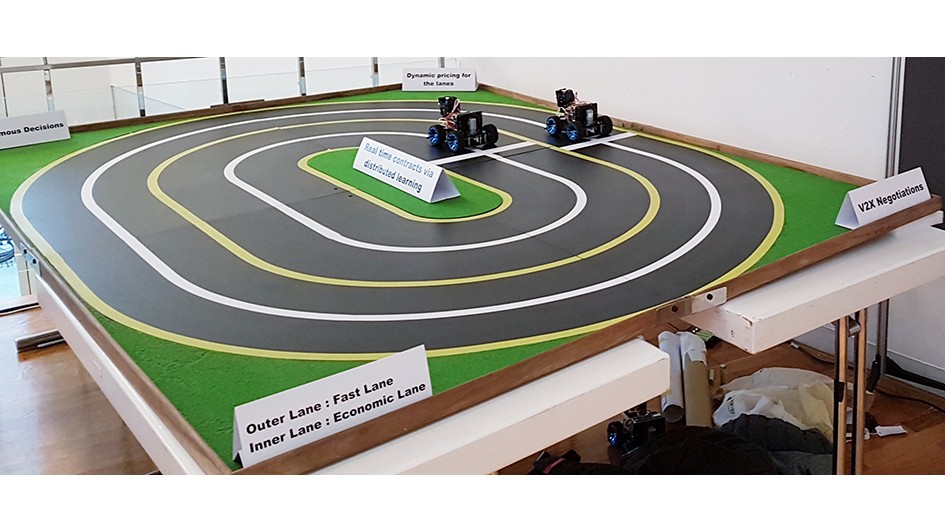}}
\caption{Snapshot of the Smart Toll Transaction Demonstrator.}
\label{fig}
\end{figure}

The demonstrator includes 6 Adeept 4WD Smart Car Kit for Raspberry Pi PiCar-B \cite{m10},
manually created roads with two lanes (Fast lane and economic lane) and 2 toll's, one on each of the lane. Each of the cars and tolls is represented as a node in the blockchain network. The agents and the tolls communicate with each other using the peer-peer protocol provided by blockchain network. Figure \ref{fig} shows a snapshot of the physical demo. 
The agent in the car understands the preferences set by the user and on the way negotiates for a faster lane or a economic lane with the Tolls. We have used python language to implement car and toll agents. Every time the car encounters the toll it negotiates with the toll and based on its preference changes the lane and enters into a negotiation contract. 
The transaction between the Toll and the Car has been captured in the quorum blockchain and the settlements between the car and toll happen instantaneously. The toll agents and the car agents wallet are also updated accordingly. 

\begin{figure*}[h]
\centerline{\includegraphics[width=0.9\textwidth]{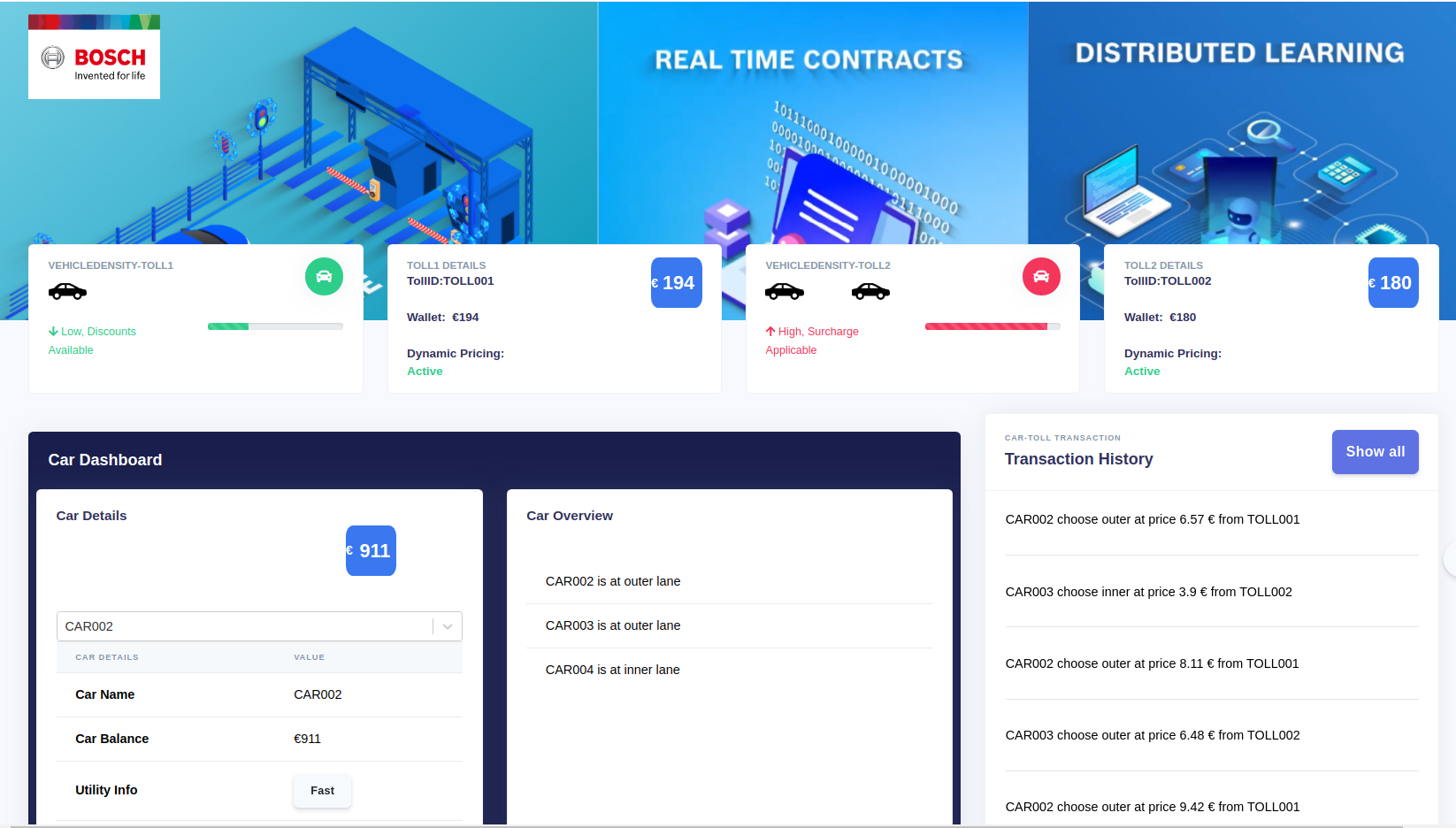}}
\caption{User Interface}
\label{fig4}
\end{figure*}

A well-structured user interface for the demonstrator was built as shown in figure \ref{fig4} using React JS which includes the vehicle details like the wallet balance, vehicle id, current lane ( fast/economic lane), toll details like the toll id, wallet balance, vehicle density in each lane and also the transaction history between the toll agents and the vehicle agents.

\begin{figure}[!h]
\centerline{\includegraphics[width=80mm,height=60mm]{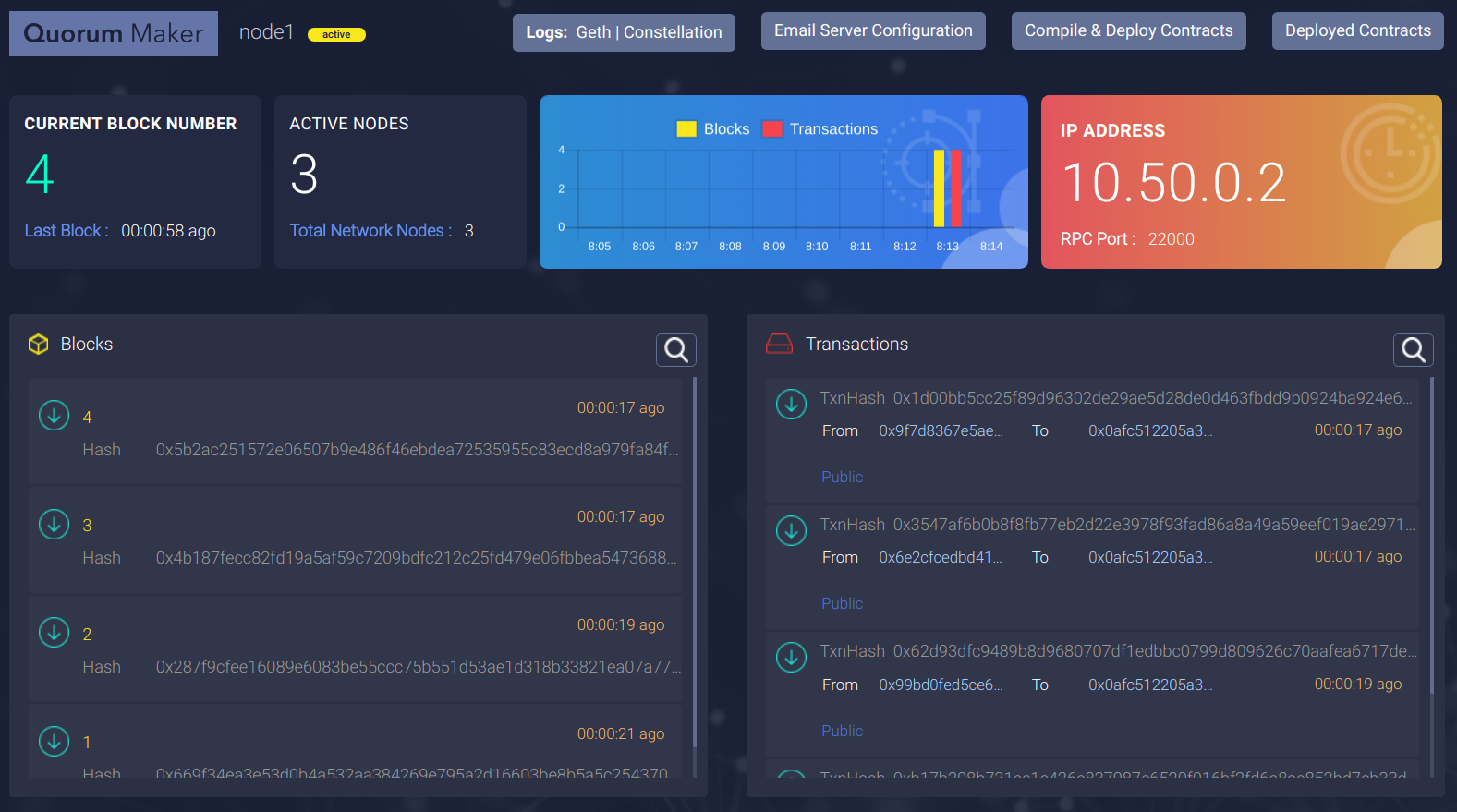}}
\caption{Quorum Blockchain Explorer}
\label{fig3}
\end{figure}

We also used the concept of smart contracts, written in solidity, to transfer ether in quorum blockchain. We implemented the quorum blockchain network using quorum-maker \cite{m101} which has an in built blockchain explorer. We could see the transactions and blocks updated, as shown in figure \ref{fig3}, when a transaction happens between vehicle and toll agents. The contract was deployed using web3.js \cite{m102} interface and was used to execute the transaction of transferring ether between vehicle agent and toll agent. 
\section{Analysis}
Our paper focuses on leveraging blockchain technology, IoT and multi-agent systems in the field of Smart Toll Transactions. The convergence of these technologies was able to bring in the features like data confidentiality, security, privacy, incentivization, collective intelligence, etc., Table \ref{tab:table-name} shows a brief overview of the features which were supported by its corresponding technology.

\begin{table}[h]
\caption{\label{tab:table-name}Values/Assets v/s Technology Used}
\begin{tabular}{|l|c|c|c|}
\hline
\multirow{2}{*}{\textbf{Values/Assets}} & \multicolumn{3}{c|}{\textbf{Technology}}                                          \\ \cline{2-4} 
                                        & IoT                       & Blockchain                & Multi Agent Systems       \\ \hline
Environmental Sensing                   & \checkmark &                           &                           \\ \hline
Data Confidentiality                    &                           & \checkmark &                           \\ \hline
Security                                &                           & \checkmark &                           \\ \hline
Privacy                                 &                           & \checkmark &                           \\ \hline
Fault Tolerance                         &                           & \checkmark &                           \\ \hline
Immutability                            &                           & \checkmark &                           \\ \hline
Incentivization                         &                           &                           & \checkmark \\ \hline
Collective Intelligence                 &                           &                           & \checkmark \\ \hline
Distributed Learning                    &                           &                           & \checkmark \\ \hline
Autonomous decision making              &                           &                           & \checkmark \\ \hline
\end{tabular}
\end{table}

\section{Summary}
With "things" around us getting connected and "things" getting enabled with intelligence, the possibilities of machine economy are endless. This paper focused on solving key problems faced in today's toll transaction using IoT, blockchain and multi-agent systems. The implementation helped us to prove that the confluence of these technologies enables new business models by leveraging IoT data monetization, automation of business rules using smart contracts and incentivization using multi-agent systems. We foresee the applicability of these technologies in many of the domains such as smart cities, smart grids, smart transportation etc.,
\section*{Acknowledgment}
We want to thank Sri Krishnan V, Mohan B V, Manojkumar Parmar, Saha Dilip from Robert Bosch Engineering and Business Solutions Private Limited, India, for their valuable comments, contributions and continued support to the project. We also thank Chatterjee Sweta, Phadke Amit Vijay, Devasthali Rajeev S  for the support of implementation and their feedback as well. We are grateful to all expert for participating in interviews and providing us with their valuable insights and informed opinions ensuring completeness of our solution implementation.

\end{document}